\documentclass[twocolumn]{aastex631}

\usepackage{hyperref}
\usepackage{amsmath}
\usepackage{subfigure}
\usepackage{multirow}
\usepackage [english]{babel}
\usepackage [autostyle, english = american]{csquotes}
\usepackage[symbol]{footmisc}

\MakeOuterQuote{"}

\usepackage{natbib}

\shorttitle{Simulating Protostellar Variability}
\shortauthors{R. R. Lee}

\graphicspath{{./}{figures/}}
\usepackage{graphicx}

\begin{document}
\title{Investigating the Role of Protostellar Variability with PRIMA Using Monte Carlo Simulations}

\author[0000-0002-7482-5078]{Rachel R. Lee}
\affiliation{Department of Physics, University of Connecticut, 196A Auditorium Road, Unit 3046, Storrs, CT 06269, USA}

\author[0000-0002-6073-9320]{Cara Battersby}
\affiliation{Department of Physics, University of Connecticut, 196A Auditorium Road, Unit 3046, Storrs, CT 06269, USA}

\author[0000-0002-6946-6787]{Aleksandra Kuznetsova}
\affiliation{Department of Physics, University of Connecticut, 196A Auditorium Road, Unit 3046, Storrs, CT 06269, USA}

\author[0000-0002-6773-459X]{Doug Johnstone}
\affiliation{NRC Herzberg Astronomy and Astrophysics, 5071 West Saanich Rd, Victoria, BC V9E 2E7, Canada}
\affiliation{Department of Physics and Astronomy, University of Victoria, 3800 Finnerty Rd, Victoria, BC V8P 5C2, Canada}

\author[0000-0002-3747-2496]{William J. Fischer}
\altaffiliation{Our beloved colleague, Will Fischer, passed away on April 16, 2024: \href{https://www.mariettatimes.com/obituaries/2024/04/william-jack-fischer/}{https://www.mariettatimes.com/obituaries/2024/04/william-jack-fischer/}.}
\affiliation{Space Telescope Science Institute, 3700 San Martin Dr, Baltimore, MD 21218, USA}

\author[0000-0002-1700-090X]{Henrik Beuther}
\affiliation{Max Planck Institute for Astronomy, Königstuhl 17, D-69117, Heidelberg, Germany}

\author[0000-0002-9017-3663]{Yasuhiro Hasegawa}
\affiliation{Jet Propulsion Laboratory, California Institute of Technology, Pasadena, CA 91109, USA}

\author[0000-0003-2248-6032]{Marta Sewi{\l}o}
\affiliation{Exoplanets and Stellar Astrophysics Laboratory, NASA Goddard Space Flight Center, Greenbelt, MD 20771, USA}
\affiliation{Department of Astronomy, University of Maryland, College Park, MD 20742, USA}
\affiliation{Center for Research and Exploration in Space Science and Technology, NASA Goddard Space Flight Center, Greenbelt, MD 20771, USA} 

\begin{abstract}
Evidence suggests that protostellar outbursts likely play a critical role in the stellar mass assembly process, but the extent of this contribution is not well understood. Using the proposed observing program of PRIMA, a concept far-IR observatory \citep[PRIMA GO Case \#43 in][]{Moullet2023}, we examine the probe's ability to unambiguously determine whether or not variable accretion events dominate the stellar mass assembly process ($M_{\rm burst}\geq0.5M_{*}$). To do this, we construct multiple protostellar ensembles using Herschel 70$\mu$m flux data and evolve them using a toy Monte Carlo simulation through steady-state and high magnitude accretion events. Ensembles are observed at various epochs in the evolution process to conclude how many large amplitude outbursts are observationally recoverable during the proposed program. Based on our synthetic observations and our simulation specifications, we determine that observing a protostellar ensemble of at least 2000 protostars using PRIMA's proposed program is sufficient for determining the importance of protostellar outbursts in the stellar mass assembly process. 

\end{abstract}


\section{Introduction} \label{sec:intro}
Stars initially gain their mass during the gravitational collapse of over-densities within a giant molecular cloud (GMC), a process in which both protostellar cores and disks are formed. As the system evolves, the protostar will continue to gain mass through the accretion of material from the surrounding disk and natal envelope. Previous work \citep{Appenzeller1972, Larson1973, Shu1977,Shu1987} has been done to understand this process using analytic models of gravitational collapse, many of which simulate this disk accretion as a steady flow of matter onto the forming protostar.

Models using this assumption of completely steady-state accretion are unable to correctly predict observed stellar mass functions using accretion rates calculated from measured protostellar luminosities \citep{Kenyon1990,Evans2009,Dunham2014,Dunham2015}. A solution to this luminosity problem, and more recently the luminosity spread problem, is the idea that mass accretion rates must be time-dependent, a theory that has been observationally reinforced by \citet{Fischer2017}. Combining time-dependent accretion rates with previous periods of rapid mass accretion may present a viable solution to this discrepancy. 

Chemical signatures and physical structures of protostellar systems indirectly point to previous epochs of these enhanced accretion rates onto protostars. For example, ice sublimation of CO, CO$_{2}$, and H$_{2}$O has been observed at radii too large to be associated with the observed protostellar luminosity, indicating a prior period of increased protostellar temperature, and thus disk temperature \citep{Kim2012,Frimann2017,Jorgensen2020}. Additionally, the presence of clumps and shocks present in protostellar jets and outflows \citep{Plunkett2015,Ray2023} have been argued to be indicators of changing mass accretion rates as a consequence of protostellar outbursts originating from mass accretion from the inner disk \citep[e.g.][]{Reipurth1989,Kim2024}.

The causal variability associated with the aforementioned signatures and structures have been observed in many systems and across all mass ranges \citep{CarattioGaratti2017}. Indeed, optical surveys have discovered luminosity increases of $\sim5$ mag  (100x luminosity) with durations of a few months to years, known as EX Lup events (EXors) \citep{Herbig1989,Aspin2010}, or on scales of years to decades, known as FU Ori events (FUors) \citep{Herbig1966,Hartmann1985}. More recently, an intermediary type of burst between FUors and EXors, known as V1647 Ori events, have been observed \citep{Acosta-Pulido2007,Ninan2013}. These events have burst amplitudes of $\sim4$ mag (40x luminosity) and durations of a few months to a few years. The details of these burst types are discussed in the review by \citet{Fischer2023}. 

These large scale outbursts have been observed across numerous wavelength regimes and star forming regions; specifically in the sub-millimeter \citep{Herczeg2017,Mairs2024}, mid-IR \citep{Cody2014, Park2021,Lee2024}, near-IR \citep{Acosta-Pulido2007,LeGouellec2024}, and optical \citep{Cody2014}. While these measurements have been able to place constraints on wavelength dependent outburst durations and amplitudes, the accretion luminosity of outbursts is highly uncertain for deeply embedded protostars at their earliest phase. Thus, we have not been able to adequately constrain the number of protostars that undergo which outbursts and for how long. 

The literature on the frequency and nature of protostellar variability is reviewed in \citet{Fischer2023} with a scientific consensus that protostellar bursts are common, but there remain many open questions about their incidence rates and durations. We know these events can present as flux increases ranging from 1x to 100x and can have durations of days to decades. While the shorter, smaller outbursts are more common, the larger, longer bursts that result in a high influx of material onto the protostar, are exceedingly rare. Due to the rarity of large outbursts, their importance in the stellar mass assembly process is currently quite uncertain \citep[see for example the timescales and frequency discussion by][]{ContrerasPena2025}.

\citet{Fischer2024} identified the far-IR observations as ideal for tracing mass accretion of embedded protostars as the far-IR does not suffer from extinction events and is at the peak of the spectral energy distributions (SEDs) emitted by these sources. Additionally, the far-IR probes the heating of the outer disk and inner envelope, which is roughly uniform regardless of viewing angle, allowing for geometry-independent measurements. The far-IR observations of protostars will be enabled by the Probe Far-Infrared Mission for Astrophysics (PRIMA), a proposed cryogenically cooled FIR probe \citep{Glenn2025} planned for the 2030s. PRIMA will carry two science instruments designed for ultra-high sensitivity imaging and spectroscopic studies in the 24-235$\mu$m wavelength range. Battersby et al. (\citeyear{Moullet2023}) argue that PRIMA is ideal for studying and monitoring the effects of this variability in the context of mass assembly of protostars. 

In this paper, we simulate synthetic observations of protostellar ensembles in which ensemble members evolve via a combination of steady-state and variable accretion. We investigate the effects of different burst types, number of burst types, and ensemble size on the ability to observationally constrain the contribution of protostellar outbursts on the stellar mass assembly process using the PRIMA observing program proposed by Battersby et al. \citep[PRIMA GO Case \#43 in][]{Moullet2023}. We test a variety of accretion events: mimicking FU Ori events, V1647 Ori events, and EX Lup events as well as routine variability seen at shorter (accretion bursters) and longer wavelengths (protostellar outbursts in the IR); hereafter RV-Short and RV-Long, respectively. In Section~\ref{sec:methods} we discuss our methodology of simulating a protostellar ensemble and evolving it with time. In Section~\ref{sec:results} we present the observed trends obtained from our simulations and discuss their takeaways in Section~\ref{sec:discussion}.

\section{Methods} \label{sec:methods}
\begin{figure*}
    \centering
    \includegraphics[width=0.98\textwidth]{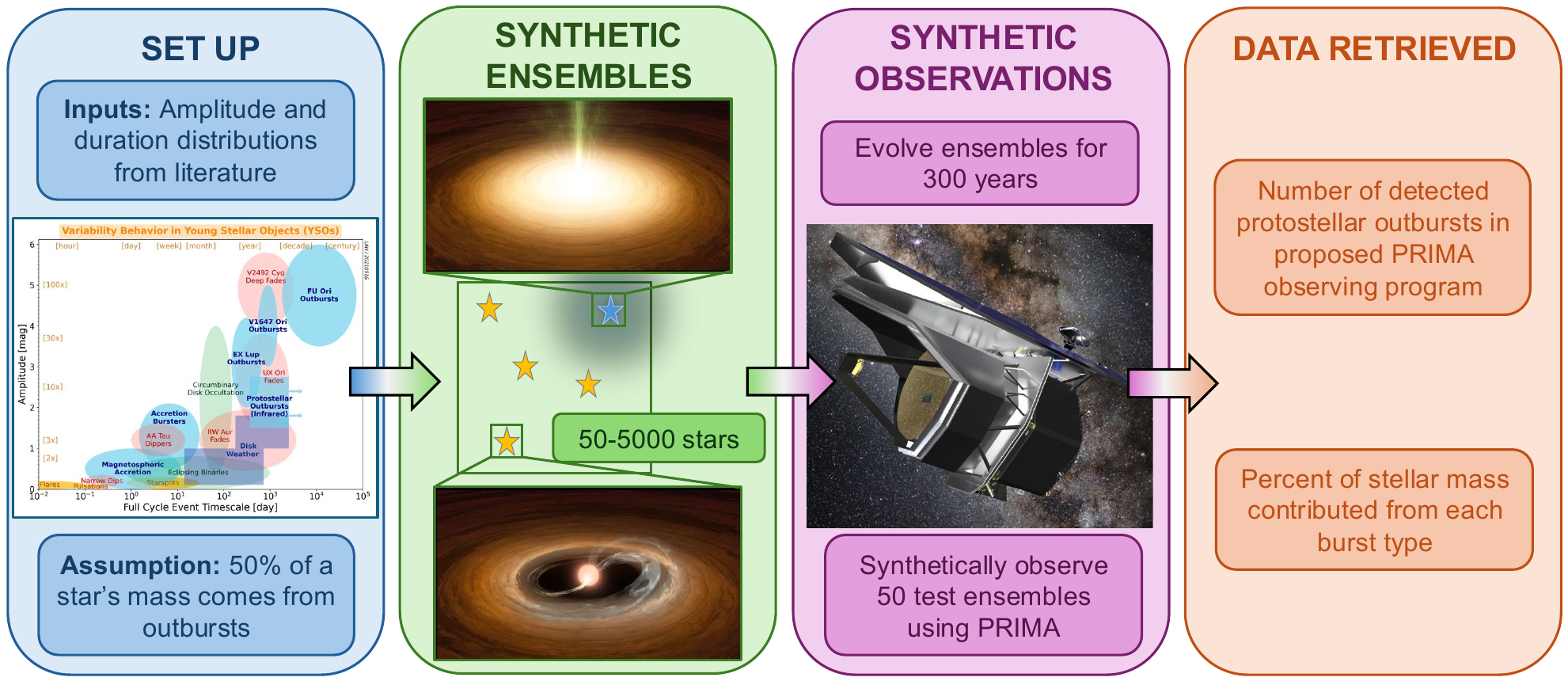}
    \caption{A schematic illustrating the steps used in this simulation. In the first panel, Monte Carlo simulation parameters are set using outburst amplitude and duration from ~\citet{Fischer2023} and assuming outbursts contribute 50\% of a star's mass. In situations in which multiple outbursts types are significant mass contributors, we assume each outburst type is equally important. In the second panel, synthetic ensembles are created by assigning 70$\mu$m Herschel flux measurements from \citet{Furlan2016} to each source in the ensemble, where ensembles can have between 50 and 5000 protostars. In panel three, 50 ensembles are evolved for 300 years and synthetically observed using the proposed PRIMA program (Battersby et al. \citeyear{Moullet2023}). In panel four, data retrieval and analysis is conducted on the number of detected outbursts by PRIMA and the associated mass contribution is calculated. Credit: Protostellar Images: T. Pyle (Caltech/IPAC); Spacecraft: NASA/JPL-Caltech; Background: ESO/S. Brunier
    \label{fig:flow_chart}}
\end{figure*}

In order to determine the impact of high-magnitude accretion bursts on the overall mass assembly process, survey design must account for the frequency and duration of these bursts. In this paper, we examine the observability of burst behavior with synthetic observations of ensembles of different sizes and different burst type allowances. We simulate the variability behavior with Monte Carlo simulations, sampling the luminosities and durations to match observed relationships. A diagram of the steps taken from simulation to synthetic observation in shown in Figure~\ref{fig:flow_chart}.

In these simulations, we have adopted a criterion for an outburst-dominant mode of mass assembly to be one in which $\geq$50\% of a star's mass comes from accretion during outburst events as described in Section~\ref{sec:case1}.

\subsection{Simulation Parameters and Set Up} \label{sec:sim_params}
For each ensemble, simulations were initialized with a population of co-eval protostars, assuming that all protostars in a single ensemble are at an early stage of evolution. Each protostar is assigned a luminosity sampled from the 70$\mu$m flux data of Class 0, Class I, and flat spectrum protostars taken with the Herschel Space Observatory \citep{Pilbratt2010} for the Herschel Orion Protostar Survey \citep[HOPS; PI: S. T. Megeath,][]{Furlan2016}. Each ensemble is evolved for 300 years via a combination of steady-state accretion and large-magnitude outbursts. To evolve ensemble members, we employed a Monte Carlo simulation to dictate which ensemble members undergo outbursts and when. Protostars in the simulation are evolved in time every two weeks, as this timescale corresponds to the duration of RV-Short bursts, the shortest burst type considered. It is assumed that each source has an equal probability of undergoing a specified type of accretion event. This probability is calculated using an assumed total mass contribution, burst duration, and burst amplitude, as detailed in Section~\ref{sec:burst_probs}. We note that each protostar is limited to one accretion burst at a time.

\begin{figure}
    \centering
    \includegraphics[width=0.47\textwidth]{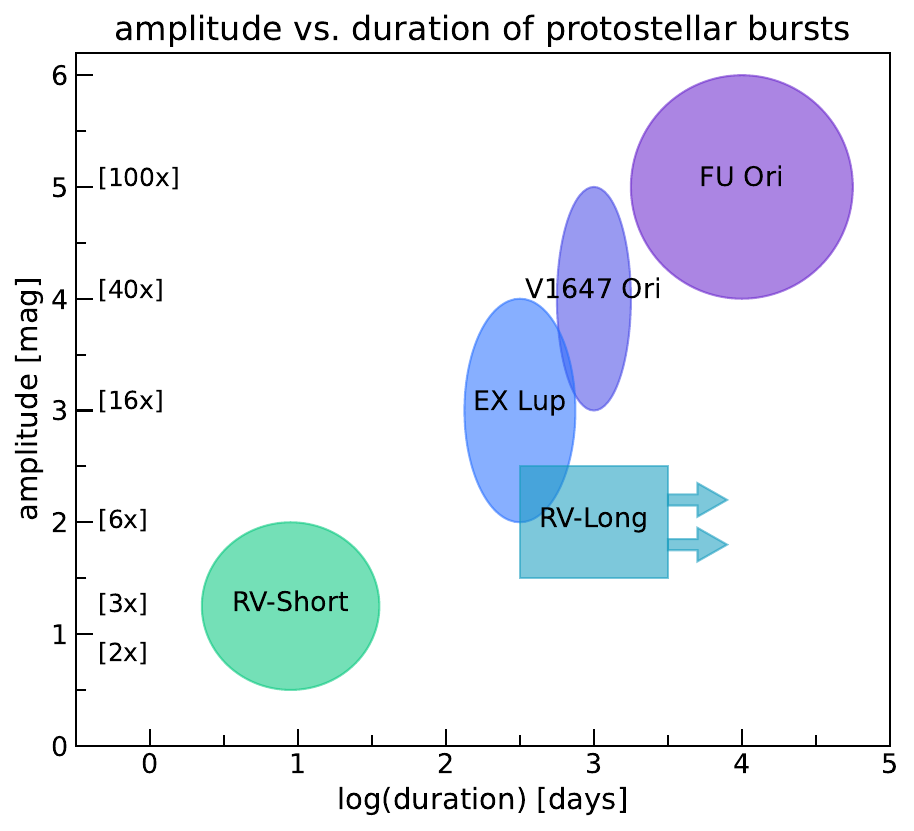}
    \caption{The duration-amplitude relationship of accretion outbursts considered in this paper: FU Ori, V1647 Ori, Ex Lup, and routine variability events (RV) of long and short duration. Central values and ranges of the different burst types (shaded regions) were obtained from \citet{Fischer2023}. In our simulations, only the central values of these regions are used (see Table~\ref{tab:burst_properties} for quantities).
    \label{fig:dur_mag_better}}
\end{figure}

Our accretion outburst duration-amplitude relationship is adopted from the range of values presented in the review by \citet{Fischer2023}. Figure~\ref{fig:dur_mag_better} shows an adapted version of Figure 3 in the review, in which we present only the outburst types considered in our investigation. We show the observationally ascertained ranges for large outburst durations and magnitudes, but use only the central values of these ranges, which can be found in Table~\ref{tab:burst_properties}. We do not consider routine variability, such as magnetospheric accretion, in our profile, as such variations are typically short period (hours to days in optical/near-IR) and low amplitude (1-2 mag) so as to have little effect on the total mass assembly accrued through a single burst. 

\begin{figure*}
    \centering
    \includegraphics[width=0.99\textwidth]{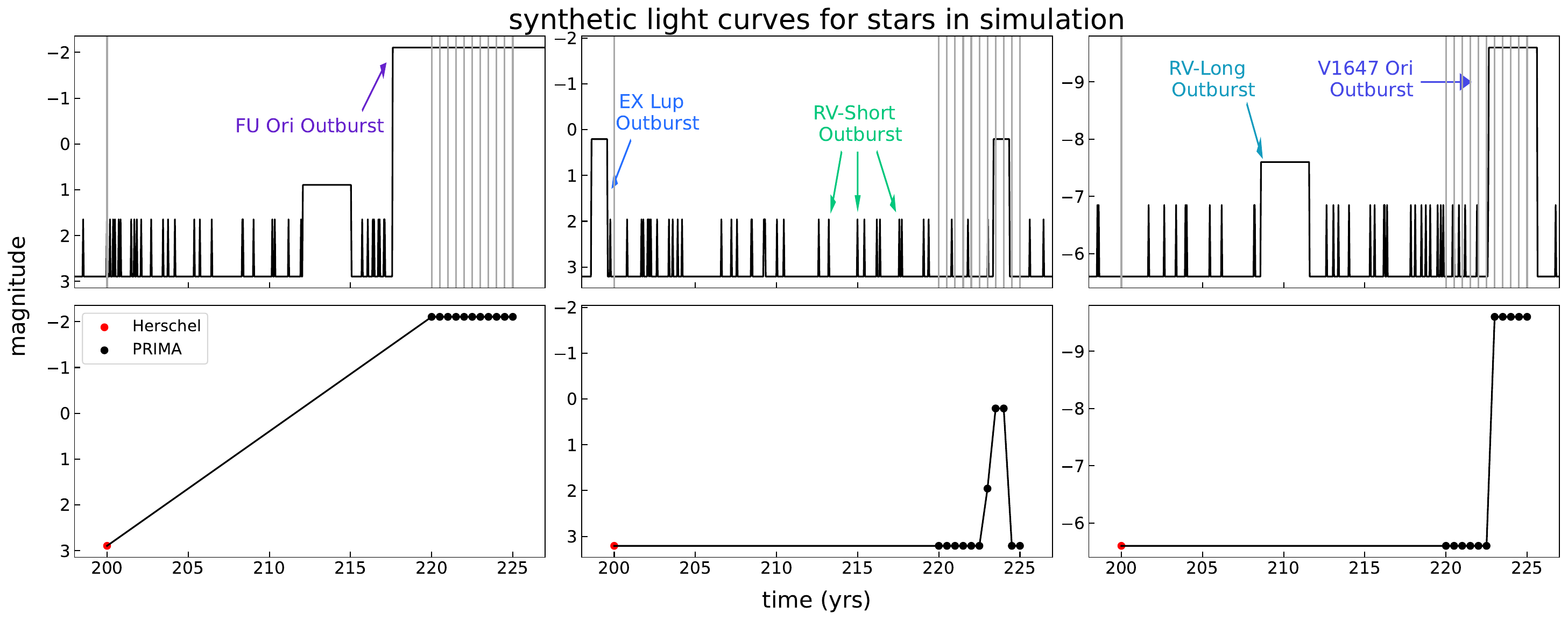}
    \caption{(Top:) Example light curves of our simulated data showing the true behavior of three protostars within our ensemble over the full 25 year observing program assuming all five burst types can occur. The grey lines indicate the timesteps in which an observation occurs. It can be seen that all burst types can be detected by the end of the program where the medium duration bursts are much more frequently and more easily observed than the larger, longer and smaller, shorter bursts. (Bottom:) The behavior of the same three protostars as observed using the proposed PRIMA observing program. Observations taken by PRIMA are displayed in black and observations taken by Herschel are displayed in red. The observing cadence blurs the true behavior of the protostars, as observed in the second panel of this row in which an RV-Short outburst appears to be associated with the brightening of an EX Lup event. 
    \label{fig:example_light_curves}}
\end{figure*}

Additionally, while observed lightcurves of outbursting objects generally exhibit time variability as well as systematic dimming and brightening \citep{ContrerasPena2025}, for simplicity, we adopt a tophat profile to model enhanced mass accretion during outburst. We present three example lightcurves in Figure~\ref{fig:example_light_curves}, in which we include the lightcurves displaying the true behavior of each protostar over the full 25 year program in the top panel, and the observational lightcurves associated with PRIMA observations in the bottom panel (observing program described further in Section~\ref{sec:PRIMA_program}). We distinguish the PRIMA observations in black and the baseline Herschel observation in red. 

In this investigation, we focused on two distinct cases: one in which mass assembly is dominated by a single mode of burst-like accretion and one in which multiple burst types can occur. In the first case, we assume that only FU Ori outbursts are significant mass contributors to the total mass of the star ($M_{\rm FUor} \geq 0.5M_{*}$). Due to the infrequency with which FU Ori outbursts are observed, an ensemble must have a large number of protostars in order for a sufficient number of outbursts to be observed. Thus, case one is used to quantify the minimum number of protostars an ensemble must have in order for PRIMA to detect a significant number of large outbursts. 

For the second case, we assumed multiple burst types are significant mass contributors, where each burst type is equally important. Adopting the nominal ensemble size informed by the results of the first case, for the second we introduce additional burst types in order to determine if the proposed PRIMA program is sufficient for determining the main contributors in the stellar mass assembly process. Simulations for each case are run for 50 realizations, producing 50 synthetic ensembles. 

\subsubsection{Case One: FU Ori Only}\label{sec:case1}
We constructed six ensembles of different sizes ranging from 50 to 5000 protostars and adopted the assumption that FU Ori outbursts are the only significant mass contributor, outside of routine accretion. Due to the rarity of FU Ori outbursts, these assumptions define an observationally limiting case. Ensembles are constructed by assigning sources a flux value from the PACS 70 $\mu$m flux data of Class 0, Class I, and flat spectrum protostars taken with the Herschel Space Observatory \citep{Pilbratt2010} for the Herschel Orion Protostar Survey (HOPS; PI: S. T. Megeath) \citep{Furlan2016}. 

We quantify the boundary between bursts being a dominant factor in the stellar mass assembly process to be when 50\% of a star's mass is accumulated during accretion bursts, $\Delta M_{\rm burst}$, over a time $\tau_{\rm burst}$. The other 50\% is accumulated through steady state (SS) accretion, $\Delta M_{\rm SS}$, over a time period of $\tau_{\rm SS}$. Under this boundary condition, 

\begin{equation}{\label{eq:SS_to_FUOri_Mass}}
    \Delta M_{\rm SS} = \Delta M_{\rm burst} \\
\end{equation}

where

\begin{eqnarray} \label{eq:mass_accumulation_eqn_array}
    \Delta M_{\rm SS} & = & \dot{M}_{\rm SS}\tau_{\rm SS} \\ 
    \Delta M_{\rm burst} & = & \dot{M}_{\rm burst}\tau_{\rm burst}
\end{eqnarray}

If we assume mass accretion rates scale with burst amplitude, $A_{\rm burst}$, we can assume $\dot{M}_{\rm burst} = \dot{M}_{\rm SS}A_{\rm burst}$. Adopting a burst amplitude of 100x for FU Ori outbursts and combining Equations~\ref{eq:SS_to_FUOri_Mass} and ~\ref{eq:mass_accumulation_eqn_array} above, the ratio of time spent in outburst to the time spent in steady state is: 

\begin{equation}{\label{eq:SS_to_FUOri_time}}
    \frac{\tau_{\rm burst}}{\tau_{\rm SS}} =  \frac{\dot{M}_{\rm SS}}{A_{\rm burst}\dot{M}_{\rm SS}} = \frac{1}{100}\\
\end{equation}

By assuming that the protostellar lifetime is the sum of time spent in outburst and the time spent undergoing steady state accretion ($\tau_{\rm lt} = \tau_{\rm SS} + \tau_{\rm burst}$), if FU Ori outbursts contribute 50\% or more of a star's total mass, the star must spend $\sim$1\% or more of its protostellar lifetime in an FU Ori state. We can then expand upon this and say that if FU Ori outbursts dominate mass assembly and any protostar can undergo a burst, then on average, 1\% of an ensemble's protostars must be undergoing an outburst at any given observation. More about mass accumulation is discussed in Section~\ref{sec:burst_probs}.

\subsubsection{Case Two: Multiple Burst Types}\label{sec:case2}
In case two, we created four different ensembles with various burst type combinations, once again constructed using flux values from the Herschel PACS 70 $\mu$m flux data. These four types of ensembles had between two and five burst types acting as contributors to the stellar mass assembly process. Each model in the multiple burst case corresponds to the N-least probable burst types, where N is the total number of burst types tested. For example, the first set of ensembles was simulated such that only FU Ori and V1647 Ori outbursts are major contributors, the second set was simulated such that FU Ori, V1647 Ori, and EX Lup outbursts were major contributors, etc. In each simulation, all burst types included are assumed to be equally important, i.e. 50\% of a star's overall mass comes from equal contributions of mass across all outburst types allowed to occur. We note that while RV-Long bursts have higher probabilities than RV-Short, we limit their inclusion to one set of models, given their departure from the period-luminosity trend in seen in Figure \ref{fig:dur_mag_better}.

\subsubsection{Calculating Input Burst Probabilities}\label{sec:burst_probs}
If a star gains its mass through a combination of steady-state accretion at some rate $\dot{M}_{\rm SS}$ and high-amplitude, $A_{i}$, outburst events with mass accretion rates defined as $\dot{M}_{i} = A_{i}\dot{M}_{\rm SS}$, the total mass accumulated, $M_{*}$, can be defined as: 

\begin{equation}{\label{eq:star_accumulation}}
    M_{*} = \dot{M}_{\rm SS}\tau_{\rm SS} + \sum_{j}^{N}\dot{M}_{\rm SS}A_{j}\tau_{j}, \\
\end{equation}
where $\tau_{\rm SS}$ is the time spent in a quiescent state, N is the number of burst types that can occur, and $A_j$ and $\tau_{j}$ are the corresponding amplitudes and time spent in outburst of burst type $j$. The summation over $j$ then represents the total mass accumulated across all outbursts. The fraction of a star's total mass accumulated from a specific burst type $i$ is then:

\begin{equation}{\label{eq:burst_mass_frac_duration}}
    f_{i,m} = \frac{M_{i}}{M_{*}} = \frac{\dot{M}_{\rm SS}A_{i}\tau_{i}}{\dot{M}_{\rm SS}\tau_{\rm SS} + \sum_{j}^{N}\dot{M}_{\rm SS}A_{j}\tau_{j}}.
\end{equation}

If for each burst type, we define the fraction of time a protostar spends in outburst relative to its total lifetime $\tau_{\rm lt} = \tau_{\rm SS}+\sum_{j}^{N}\tau_{j}$ to be $f_{i,t} = \tau_{i}/\tau_{\rm lt}$, the fractional mass associated with that burst type can be expressed in terms of the fractional time spent in outburst:  

\begin{equation}{\label{eq:burst_mass_frac_time}}
    f_{i,m} = \frac{A_{i}f_{i,t}}{(1-\sum_{j}^{N}f_{j,t}) + \sum_{j}^{N}A_{j}f_{j,t}}.
\end{equation}

For a specific burst type, the probability of that type of burst occurring per year is then the ratio of the fractional time spent in outburst, $f_{i,t}$, to the typical burst duration, $t_{i}$: 

\begin{equation}{\label{eq:burst_prob}}
    p_{i}=\frac{f_{i,t}}{t_{i}}.
\end{equation}

Note that $t_{i}$ is different from $\tau_{i}$ as the former represents the outburst duration and the latter represents the total time spent in an outburst of type $i$. 
An example of the calculation used for both case one and two can be seen in Sections~\ref{sec:calculation_case_one} and \ref{sec:calculation_case_two}, respectively. The corresponding burst amplitude, duration, and probability for the varying burst types investigated in this paper can be seen in Table~\ref{tab:burst_properties}.  

We briefly note that probabilities do not scale linearly with the number of outbursts allowed in a simulation, as may be expected from simple mathematical assumptions. This is a result of stellar mass being calculated from total protostellar lifetime explicitly as opposed to assuming $\tau_{\rm lt} \sim \tau_{\rm SS}$. We adopt this treatment with the understanding that as the number of burst types allowed in a simulation increases, the time spent in steady state becomes less representative of protostellar lifetime, a discrepancy that becomes exacerbated once smaller burst types are considered. 

\begin{deluxetable*}{cccccccc}
\tablecaption{Accretion Burst Types and Properties \label{tab:burst_properties}}
\tablehead{\colhead{Burst Type} & \colhead{Amplitude}& \colhead{Duration} &
\multicolumn{5}{c}{Burst Probability if 50\% of Stellar Mass Comes From Bursts}\\
\colhead{} & \colhead{} & \colhead{(yrs)} & \multicolumn{5}{c}{(burst yr$^{-1}$)}\\
\cline{4-8}
\colhead{} & \colhead{} &
\colhead{} & \colhead{1 burst} & \colhead{2 bursts} & \colhead{3 bursts} & \colhead{4 bursts} & \colhead{5 bursts}}
\startdata
FU Ori &  100 & 30 & $3.30 \cdot 10^{-4}$ & $1.64 \cdot 10^{-4}$ & $1.08 \cdot 10^{-4}$ & $7.52 \cdot 10^{-5}$ & $5.96 \cdot 10^{-5}$ \\
V1647 Ori &  40 & 3 & - & $4.10 \cdot 10^{-3}$ & $2.69 \cdot 10^{-3}$ & $1.88 \cdot 10^{-3}$ & $1.49 \cdot 10^{-3}$ \\
EX Lup &  16 & 1 & - & - & 0.0202 & 0.0141 & 0.0111 \\
RV-Short &  3 & 0.04 & - & - & - & 1.881 & 1.489 \\
RV-Long &  6 & 3 & - & - & - & - & 0.0099
\enddata
\tablecomments{Properties for each model in this paper: burst amplitude, duration, and probability for all burst scenarios we have investigated.}
\end{deluxetable*}

\subsection{Synthetic Observations} \label{sec:syn_obs}
\subsubsection{PRIMA Observing Program}\label{sec:PRIMA_program}
The synthetic observations used in these simulations are designed to resemble that of a PRIMA observing program proposed in GO case \#43 \citep{Moullet2023} and PRIMA GO Case \#105 (in preparation). The proposed program aims to monitor a protostellar ensembles over the full five year PRIMA lifetime, observing once every six months. Observations can readily track brightness variations due to the excellent sensitivity of PRIMA \citep{Burgarella2023,Bradford2023}. These flux values will then be compared to 2009 Herschel flux values to gauge the longer-term variations in the brightness. 

We assume that observed flux values correspond to accretion rates using the following equation for accretion luminosity \citep{Gullbring1998}:

\begin{equation}
    L_{\rm acc} = \frac{\eta\dot{M}GM_{*}}{R_{*}},
\end{equation}
where $\eta$ represents the efficiency of converting accretion energy into radiation. Thus, we take that the accretion luminosity in a burst state increases proportionally to the steady-state such that $L_{b} = A_b L_{\rm SS}$. 

In order to generate synthetic lightcurves from the simulations (see Figure~\ref{fig:example_light_curves}), we sample from a random start time, $t=0$ within the sample, avoiding 20 years at the start and end of the simulation. This buffer allows us to account for scenarios in which PRIMA observations occur during an ongoing burst of long duration, particularly FU Ori outbursts. Synthetic observations are generated for the ensemble initially at randomly sampled $t=0$ to represent the Herschel observations, then eleven additional times between $t=20-25$ years at an observing cadence of six months for the PRIMA campaign. 

\subsubsection{Burst Detections}\label{sec:burst_detections}
With the simulated light curves obtained from the synthetic observations detailed above, we determined the number and type of outbursts that would be detected during the survey lifetime. For a point in the lightcurve of a single source to count as a burst detection, we require that the observation must have an elevated amplitude relative to the minimum luminosity within the lightcurve, which we take to represent the quiescent luminosity of the protostar. In practice, we subtract the minimum luminosity to generate a quiescent subtracted light curve that maps to the outburst magnitudes.

The total number of bursts in a quiescent-subtracted lightcurve corresponds to the frequency of non-consecutive repetitions of a single amplitude value, that is, consecutive repetitions of a single amplitude represent a single burst. Outburst classifications are based on amplitude ranges of equally spaced logarithmic bins centered around values found in Table~\ref{tab:burst_properties}. To generate probability distributions we sum burst detections across all synthetic ensembles for each set of models. 

When distinguishing outbursts, we do not consider burst duration as a factor of our classifications. Thus, back-to-back outbursts or scenarios in which the minimum magnitude is not representative of the true quiescent magnitude bursts may be classified incorrectly. By design, this accurately mimics problems true observing programs would face. 

\subsubsection{Determining Mass Accumulation Percentages}\label{sec:mass_percentages}
The fractional mass accumulated from each burst type is directly proportional to the fraction of the star's lifetime spent in that outburst state, $f_{i,t}$. Due to outburst durations being comparable to PRIMA's lifetime, we make the assumption that $f_{i,t}$ is equivalent to the fraction of observations seen in outburst. This value is represented by the ratio of the number of observations taken during outburst, $n_{\rm obs,burst}$, to the total number of observations in a lightcurve, $n_{\rm obs}$. The total mass accumulation percentage for each burst type is then calculated from equation \ref{eq:burst_mass_frac_time} using the fractional observing time spent in outburst for each burst type. The distributions are generated based on results across the 50 synthetic ensembles for each model type. 

\section{Results}\label{sec:results}
Any observing program aiming to measure if stellar mass assembly is dominated by accretion events will be limited by the rarest and most powerful events, FU Ori outbursts (\S \ref{sec:results_ensemble_size}). Using only these bursts as our limiting case, we determine the minimum number of protostars \citep[observed as proposed in PRIMA GO case \#43 in][]{Moullet2023} required to definitively answer whether accretion events dominate the mass assembly process. Based on this minimum, we adopt a nominal ensemble size for models with multiple burst types to determine the proposed PRIMA observing program's capabilities for characterizing mass assembly for multiple durations and amplitudes of outburst (\S \ref{sec:multiple_bursts_results}). 

For each Monte Carlo toy model of a protostellar ensemble, we run 50 realizations. The resulting distributions presented in this section represent the probability spread of the number of outburst detections and associated mass contribution percentages for each simulation. 

\subsection{Investigating Multiple Ensemble Sizes}\label{sec:results_ensemble_size}
In our first case, we assume burst contributions are solely from FU Ori events and investigate the effect of multiple ensemble sizes on the accuracy of derived mass accretion percentages. In this scenario, as discussed in Section~\ref{sec:case1}, a protostar must be undergoing an FU Ori outburst $\sim1$\% of the total accretion time in order to contribute 50\% of a star's mass. Using the steps outlined in Section~\ref{sec:burst_probs} and assuming that if only FU Ori outbursts contribute 50\% of the star's mass (the other 50\% from steady-state accretion) and each burst lasts for 30 years, Equation~\ref{eq:burst_prob} gives a burst probability of $\sim0.00033$ bursts per year for each individual protostar, which can be applied to all ensemble sizes. For this case, synthetic ensembles have sizes of 50, 100, 500, 1000, 2000 and 5000 protostars.  

\begin{figure}
    \centering
    \includegraphics[width=0.445\textwidth]{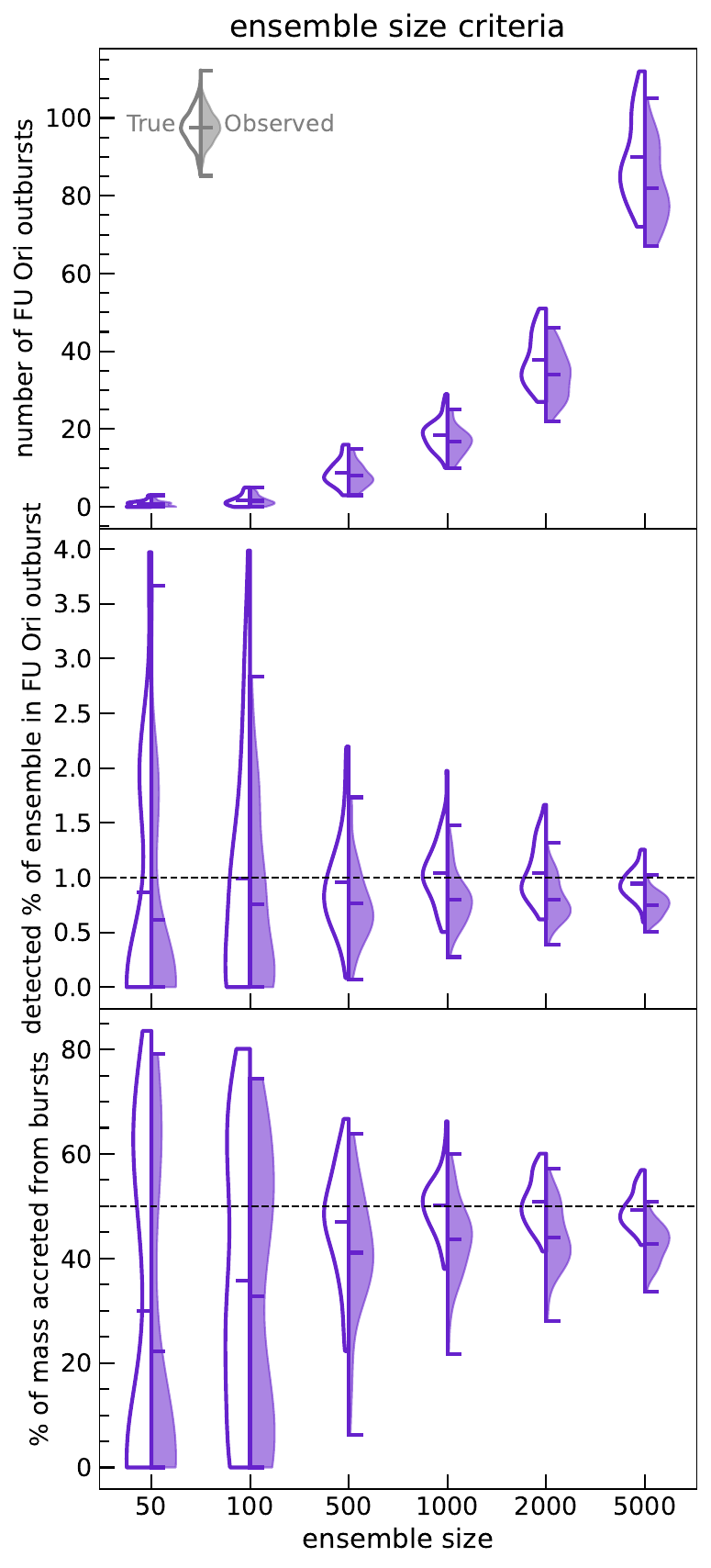}
    \caption{(Top): Total number of outbursts that occur over 25 years for different ensemble sizes. The white left hand side of each violin plot shows the distribution of bursts happening in the simulation (True), regardless of observing cadence. The shaded right hand side represents the number of bursts PRIMA is able to detect over its lifetime (Obs.) for an observing cadence of six months. (Middle): The percent of an ensemble's protostars undergoing a detected FU Ori outburst during any of the observations in the observing program. (Bottom): The stellar mass accumulation percentage contributed by FU Ori bursts under the assumption that FU Ori bursts accumulate 50\% of stellar mass, indicated with a black dashed line. 
    \label{fig:ensembleSizeChange_AllPlots}}
\end{figure}

We present the probability distributions of the number of bursts that would occur during the total length of the 25 year observing program, the percent of an ensemble undergoing an outburst during any given observation, and the associated mass accumulation percentages in Figure~\ref{fig:ensembleSizeChange_AllPlots}. In all panels we compare the true and observed distributions of these properties. True distributions associated with the full 25 year program lifetime are indicated in white on the left hand side (LHS) and observed distributions associated with PRIMA observations are indicated on the shaded right hand side (RHS). A table containing the averages of the distributions shown in this plot can be see in Table~\ref{tab:ensemble_size_properties}. 

\begin{deluxetable*}{ccccccc}
\tablecaption{True and observed properties of multiple ensemble sizes with only FU Ori outbursts occurring}
\label{tab:ensemble_size_properties}
\tablehead{\colhead{ensemble Size} & \colhead{$N_{true} (\sigma_{true})$}& \colhead{$N_{obs} (\sigma_{obs})$} & \colhead{$\%_{true} (\sigma_{true})$} & \colhead{$\%_{obs} (\sigma_{obs})$} & \colhead{$M_{true} (\sigma_{true})$} & \colhead{$M_{obs} (\sigma_{obs})$}} 
\startdata
50 &  0.8 (0.8) & 0.7 (0.8) & 0.9 (1.2) & 0.6 (0.9) & 29.9 (30.6) & 22.4 (28.8) \\
100 &  1.7 (1.5) & 1.6 (1.3) & 1.0 (1.1) & 0.8 (0.8) & 35.8 (26.4) & 32.8 (25.7) \\
500 &  8.8 (3.3) & 8.0 (3.1) & 1.0 (0.4) & 0.8 (0.4) & 47.1 (10.4) & 41.1 (12.0) \\
1000 &  18.5 (3.8) & 16.8 (3.7) & 1.0 (0.3) & 0.8 (0.3) & 50.2 (5.5) & 43.7 (7.7) \\
2000 &  37.9 (6.6) & 34.0 (6.1) & 1.0 (0.3) & 0.8 (0.2) & 50.8 (4.5) & 44.0 (6.0) \\
5000 &  89.9 (10.6) & 82.0 (9.8) & 0.9 (0.1) & 0.8 (0.1) & 49.2 (3.3) & 42.8 (3.7) 
\enddata
\tablecomments{The true versus observed properties of ensembles containing between 50-5000 protostars. Properties included are the number of outbursts occurrences (true) and detections (observed), percent of an ensemble undergoing an outburst at any given PRIMA observation, and the associated mass accumulation percentages. In parenthesis the 1$\sigma$ deviation about the mean is reported.}
\end{deluxetable*}

In the top panel of Figure~\ref{fig:ensembleSizeChange_AllPlots}, we show the true and observed number of FU Ori outbursts happening over the 25 year program. We see that the proposed observing program is able to recover, on average, $\sim$91\% of outbursts across all ensemble sizes. This recovery percentage is as expected when considering how program lifetime compares to the outburst duration and outburst detections being based upon brightening and dimming events. This means that the true number of outbursts is based upon brightening events occurring over a 55 year time period (25 year observing program and 30 year burst duration). On the other hand, the observed number of bursts is based on  50 years of evolution as there is a five year period in which a protostar will have an elevated amplitude for the entirety of the observing program, and the outburst will go undetected.

Despite similarities in burst recovery percentages, smaller ensemble sizes (50-500) show a high degree of variation, ranging from 10\% for 500 protostars to nearly 30\% for 50 protostars. Importantly, we see that as the ensemble size increases, the likelihood of detecting multiple outburst increases. For both ensemble sizes of 50 and 100 protostars, there is a 25-50\% chance no outbursts are detected at all. In larger ensemble sizes, there is at least one outburst detected in each realization.

In the middle panel of Figure~\ref{fig:ensembleSizeChange_AllPlots}, we show what percent of an ensemble's protostars is undergoing an active outburst at any given observation as a function of the total ensemble size. The ideal 1\% of an ensemble undergoing an outburst with a black dashed line. On average, outburst detections indicate 0.8\% of an ensemble undergoing an active outburst during an observation. This decreased ensemble percentage is systematically offset by $\sim0.2\%$ in all ensemble sizes with slight variation (0.25\% for an ensemble of 50 protostars to 0.19\% for an ensemble of 5000 protostars). This decreased percentage and systematic offset is a consequence of PRIMA missing five years of outburst activity, as a result we expect to miss $~20\%$ of outbursts and by extension, $~0.2\%$ of outburst during any given observation.

We also find that as ensemble size increases, the probability spread of detection percentage decreases from nearly 1\% for an ensemble of 50 protostars to 0.1\% for an ensemble of 5000 protostars. A significant narrowing in the spread occurs once an ensemble has at least 500 protostars, providing a coefficient of variation (CV) of less than 1.0. Further increasing the ensemble size to at least 1000 protostars provides a CV of $\leq0.3$. We also briefly note that all ensemble sizes, except that of 5000 protostars, observe an average ensemble percentage undergoing an outburst within $1\sigma$ of the ideal 1\%, further suggesting a systematic offset between true and observed values. 

With the number of outbursts detected during a single observation over the entire observing program, we use Equation~\ref{eq:burst_mass_frac_time} to determine the fractional mass contributed by both FU Ori outbursts and steady-state accretion. The probability distribution of the mass accumulation percentages can be seen in the bottom panel of Figure~\ref{fig:ensembleSizeChange_AllPlots}, with the desired 50\% mass accumulation rate indicated with a black dashed line. 

We find that as the ensemble size increases, the mass accumulation percentage contributed by FU Ori outbursts tends toward 43\%. This decreased mass accumulation percentage is expected as these values are a direct result of the number of protostars undergoing an outburst during any given observation. The two smallest ensembles present low average mass accumulation percentages of $<33\%$ with high degrees of variation in both the true and observed mass accumulation percentages, $\sigma\sim 26-29\%$. As the ensemble sizes increase, mass accumulation percentages increase and the spread in the probability distribution decreases to $\sigma<7\%$ once an ensemble has at least 1000 protostars.

Using the bottom panel of Figure~\ref{fig:ensembleSizeChange_AllPlots}, we compare the true mass accumulation percentages (LHS) to the observed mass accumulation percentages (RHS) to test the accuracy associated with the proposed PRIMA program. We find that as the ensemble size increases and probability distributions narrow, the proposed program's ability to recover the total mass accumulation percentage improves significantly. The proposed program is only able to recover 74\% of mass accumulation for an ensemble of 50 protostars, but once ensembles have at least 1000 sources, $\sim$90\% of the true mass accumulation is recovered by PRIMA. 

Based on our findings from Figure~\ref{fig:ensembleSizeChange_AllPlots} and further discussion presented in Section~\ref{sec:discussion_size_change}, we adopt a fiducial ensemble size requirement of 2000 protostars for the remaining simulations. We note that the assumption of 1000 protostars is also appropriate for this work, but adopt a factor of two science margin. 

\subsection{Investigating Multiple Burst Types}\label{sec:multiple_bursts_results}
Adopting a nominal ensemble size of 2000 protostars, we begin investigating case two, in which multiple outburst types contribute to the overall mass of a star. In this case, we simulate burst detections with additional contributions from a variety of burst types spanning a range of durations and amplitudes: RV-Short, RV-Long, EX Lup, and V1647 Ori (Figure \ref{fig:dur_mag_better}). In this work, we focus on the most extreme scenario in which five burst types may occur. Analysis of scenarios in which two, three, and four burst types may occur has been conducted, but left out of this investigation as the trends observed in the five burst type scenario are consistent across all simulation set ups. Probabilities for these scenarios can be found in Table~\ref{tab:burst_properties}. 

\begin{figure*}
    \centering
    \includegraphics[width=0.97\textwidth]{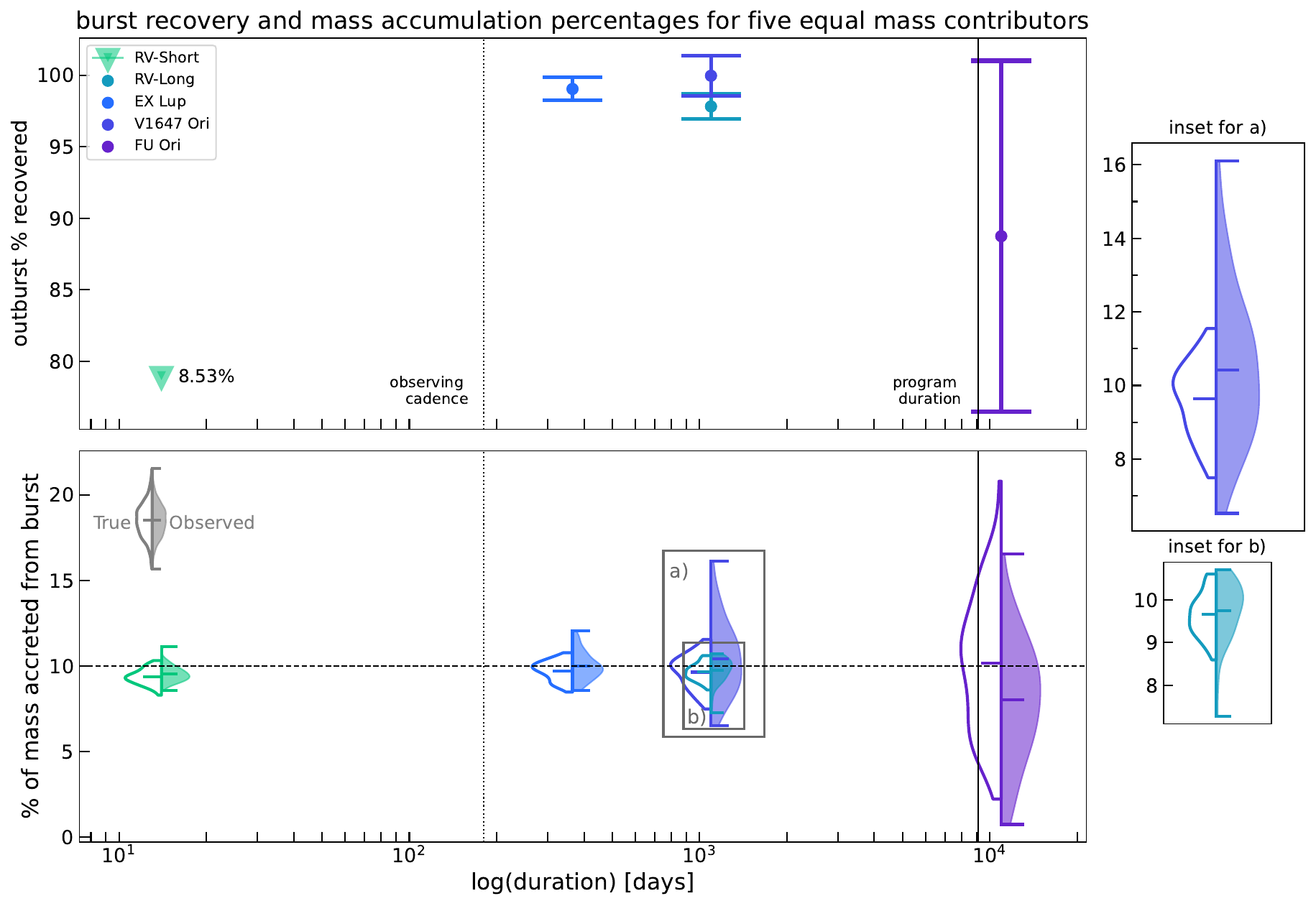}
    \caption{(Top): Percentage of outbursts recovered over the duration of the program (black solid line). Shows that recovery percentages are robust when outburst duration is within the limits of the observing cadence (black dotted line). Error bars represent $1\sigma$ deviation about the mean. RV-Short outbursts are represented with a downwards triangle as they are out of range for this plot with a burst recovery percentage of 8.53\%. 
    (Bottom): Mass accumulation percentages show that despite a significant amount of missed burst detections, recovery percentages are not the sole factor in determining mass accumulation percentages. All burst types show accumulation percentages within one sigma of the true accumulation percentages. 
    (Top/Bottom insets): Zoomed in view of the mass accumulation probability distribution of V1647 Ori outbursts (upper inset, a) and RV-long outbursts (lower inset, b). 
    \label{fig:five_burst_types_AllPlots}}
\end{figure*}

A distribution of the number of outburst detections and associated mass assembly percentages can be seen in Figure~\ref{fig:five_burst_types_AllPlots} with true and synthetically observed values presented in Table~\ref{tab:5bin_properties}. As mentioned previously, these distributions and values are based only on simulation realizations in which all five burst types are equal mass contributors.

\begin{deluxetable*}{cccccc}
\tablecaption{True and observed properties of an ensemble of 2000 stars with all five burst types occurring \label{tab:5bin_properties}}
\tablehead{\colhead{Burst Type} & \colhead{$N_{true} (\sigma_{true})$}& \colhead{$N_{obs} (\sigma_{obs})$} & \colhead{\% Recovered} & \colhead{$M_{true} (\sigma_{true})$} & \colhead{$M_{obs} (\sigma_{obs})$}} 
\startdata
FU Ori &  6.6 (2.1) & 5.8 (2.1) & 88.8 & 10.1 (4.0) & 8.0 (3.4) \\
V1647 Ori &  31.9 (5.0) & 31.9 (5.3) & 100.0 & 9.6 (1.0) & 10.4 (2.1) \\
EX Lup &  151.6 (11.9) & 150.2 (11.9) & 99.0 & 9.7 (0.6) & 10.0 (0.8) \\
RV-Short &  145243.3 (159.4) & 1238.5 (35.7) & 8.5 & 9.4 (0.5) & 9.7 (0.7) \\
RV-Long &  208.0 (11.2) & 203.5 (11.3) & 97.8 & 9.7 (0.5) & 9.5 (0.5)
\enddata
\tablecomments{The true versus observed properties of ensembles containing 2000 stars in which five outburst types contribute 50\% of a star's mass. Properties included are the number of outbursts occurrences (true) and detections (obs.), PRIMA's recovery percentage of outbursts, and the associated mass accumulation percentages. In parenthesis the 1$\sigma$ deviation about the mean is reported.}
\end{deluxetable*}

From the top panel of Figure~\ref{fig:five_burst_types_AllPlots}, we see that PRIMA is able to recover nearly all ($\sim 97-100\%$) of the medium length outbursts (V1647 Ori, EX Lup, and RV-Long) consistently across all realizations. For FU Ori outbursts, the longest and rarest outbursts, recovery rates are somewhat lower and more variable: $88.8\% \pm 12.3 \%$, an expected value as discussed in Section~\ref{sec:results_ensemble_size}. In contrast, RV-Short outbursts have very short durations with respect to the observing cadence and are much less likely to be detected within the observing window, at a rate of $8.5 \% \pm 0.2\% $. This decreased percentage is anticipated and is a consequence of RV-Short outbursts lasting for two weeks, $\sim8\%$ of the six month observing cadence.

Following methods outlined in Section~\ref{sec:mass_percentages}, we use the number of detections to calculate the observed mass contribution from each burst type to the overall stellar mass. These distributions can be seen in the bottom panel of Figure~\ref{fig:five_burst_types_AllPlots} with a comparison to the actual (input) mass contributions in the simulation, about $10\%$ (indicated with a black dashed line) from each burst sums up to half of the star's mass accumulated through outburst events. We find that all observed distributions recover 10\% of the mass within 1$\sigma$ on average. The two longest/high amplitude bursts are observed to have uncertainties on the mass fraction that are larger than those of the three other burst types (3.4\% for FU Ori and 2.0\% for V1647 Ori). It can be seen that the large spread in FU Ori mass accumulations is present both the observed and true accumulation distributions. Overall, we find that the proposed observing program can recover an average of 98.4\% of the input mass contribution, with some variation by burst type ranging from $79\%$ for FU Ori to $108\%$ for V1647 Ori.

\section{Discussion}\label{sec:discussion}
\subsection{Nominal Minimum Ensemble Size} \label{sec:discussion_size_change}
As mentioned in Section~\ref{sec:results_ensemble_size}, we adopt a nominal ensemble size of 2000 protostars for our simulations. Under limiting case assumptions ($M_{FU Ori} \geq 0.5M_{star}$), this ensemble size was selected based on three criteria: the most probable number of outbursts observed over the duration of the proposed PRIMA program, the fraction of the ensemble undergoing an outburst during any PRIMA observation, and the associated mass accumulation percentages. The final two criterion focus on having a narrow enough distribution in both the percent detected and percent mass accreted to uncover the underlying percentages. 

When examining the first selection criterion (top panel of Figure~\ref{fig:ensembleSizeChange_AllPlots}), it is clear that outburst detections are unlikely in small ensemble sizes with 50\% of realizations showing no outbursts for an ensemble size of 50 protostars and 25\% of realizations for 100 protostars. In ensembles with at least 500 sources, PRIMA consistently observes at least one outburst throughout the duration of the proposed program. Thus, the first criteria tells us an ensemble must contain 500 protostars at minimum in order for any large protostellar outbursts to be reliably observed. 

Investigating the second criterion (Figure~\ref{fig:ensembleSizeChange_AllPlots}, middle panel), we see that in the proposed PRIMA program recovers $\sim0.8\%$ of each ensemble undergoing an active outburst per observation on average with a systematic offset from the true percentage of $\sim0.2\%$. It can clearly be seen that the reliability of recovering this ensemble percentage increases significantly with increasing ensemble size, as the spread of probability distributions decreases drastically, most noticeably once an ensemble has at least 500 protostars. Combining the precision of ensemble percentage detections and the consistent offset from the true ensemble percentage, it is clear that under our adopted assumptions, 0.8\% of an ensemble being observed to be undergoing an FU Ori outburst is indicative of 50\% of a star's mass being contributed by FU Ori outbursts, no matter ensemble size. 

For our last criterion, considering ensembles containing at least 500 protostars, we then investigate the mass accumulation percentages that the proposed PRIMA program is able to recover for each ensemble size (bottom panel of Figure~\ref{fig:ensembleSizeChange_AllPlots}). It is apparent that ensembles of 1000-5000 protostars observationally recover roughly the same mass accumulation percentage, with an ensemble of 500 protostars displaying a mass accumulation percentage $\sim3\%$ lower than the larger ensembles. This ensemble size also shows a much higher uncertainty, indicating its probability distribution is not narrow enough for us to reliably ascertain mass accumulation percentages. 

For the larger ensemble sizes (1000-5000), we find all three ensembles appear to indicate roughly the same mass accumulation percentages, with a slight decrease of $\lesssim1\%$ for an ensemble of 5000 protostars, and minor differences in probability spreads. We can also see that the proposed PRIMA program detects mass accumulation percentages more accurately in an ensemble of 2000 protostars, if only marginally. With subtle differences between the observational properties of these ensembles, we conclude that an ensemble of 5000 protostars does not yield distinctive scientific knowledge over those a 1000 or 2000 protostars. Thus, we adopt a minimum ensemble size of 2000 protostars, despite limited statistical distinction, to account for a factor of two scientific margin.

\subsection{Multiple Burst Types} \label{sec:discussion_multi_bins}
Adopting a nominal ensemble size of 2000 protostars, we now investigate the effects of increasing the number of burst types in a simulation. Analysis is focused on how recoverability of protostellar outburst and the associated mass accumulation percentages are impacted. 

In the top panel of Figure~\ref{fig:five_burst_types_AllPlots}, it is clear that the proposed program's ability to detect outbursts is maximized for V1647 Ori, EX Lup, and RV-Long outbursts. This can be attributed to their durations falling between the window of the observing cadence and the program duration. Conversely, FU Ori and RV-Short outbursts have much lower recovery percentages. This is expected as both burst types have durations outside the bounds of the program lifetime and observing cadence. It is also expected that FU Ori outbursts are more readily recovered than RV-Short outbursts as the severity with which outbursts are missed is based upon the relationship between outburst duration and program related timescales. 

Despite over- and under-predictions of outbursts, observation based mass accumulations are fairly accurate (see the bottom panel of Figure~\ref{fig:five_burst_types_AllPlots}). Mass accretion percentages are within $1\sigma$ of 10\% on average for all burst types and observed percentages deviate only slightly from the true mass accumulation percentage in the simulation, with the exception of FU Ori. We also see a large spread in both the observed and true mass accumulation percentages associated with FU Ori outburst, likely due to the rarity of FU Ori events. In spite of this, the large number of other burst types the proposed program is able to detect allows us to determine mass accumulation percentages with relatively high accuracy. 

These findings indicate that the proposed PRIMA observing program is able to determine mass accumulation percentages from protostellar accretion with reasonable accuracy. The limiting sample size is set by the possibility that most of the mass is accumulated in rare but powerful FU Ori outbursts. In this case, with our present understanding of these bursts, PRIMA's observing program could determine the mass accumulation percent to a $1\sigma$ accuracy of 6.0\%. Alternatively, if there are multiple bursts that make meaningful contributions to the stellar mass assembly process, PRIMA will see plenty of the V1647 Ori, EX Lup, RV-Long, and RV-Short outbursts with an accurate mass accumulation percentage ($1\sigma$ certainty of $<2$\%) and a 3.4\% certainty in mass accumulation percentage for the rarer FU Ori outbursts. Therefore, we argue that the proposed observing program for PRIMA is sufficient to determine the role of protostellar outbursts in the stellar mass assembly process.

\subsection{Interpretation of Observed Mass Accumulation Percentage} \label{sec:discussion_interpretation}
Due to the program lifetime and observational cadence, observed mass percentages are systematically offset from true mass accumulation percentage. Using Monte Carlo simulations, the magnitude of correction between these values can be quantified. 

For an ensemble of 2000 protostars in which only FU Ori outbursts are the main contributor of a star's total mass, observed mass accumulation rates are systematically offset from the true mass accumulation rate by 6.8\% (columns six and seven in Table~\ref{tab:ensemble_size_properties}). The magnitude of this correction factor is also observed in case two, in which the observed mass accumulation of FU Ori outbursts is systematically offset from the true mass accumulation rate by 2.1\% (columns five and six in Table~\ref{tab:5bin_properties}). Looking closer at the simulation with all five burst types occurring, mass accumulation offsets for other burst types are much smaller than that of FU Ori. Associated correction factors are 0.78\% for V1647 Ori, 0.28\% for EX Lup, 0.09\% for RV-Long, and 0.15\% for RV-Short. Applying these correction factors result in all realizations presenting burst specific mass accumulation rates within $1\sigma$ of the associated true mass accumulation rates. 

\section{Conclusions}\label{sec:conclusion}
In this paper, we simulate the evolution of protostellar ensembles in which outbursts are a dominant form of mass contribution for the protostars (i.e. in which 50\% of the star's total mass is accumulated through bursts), using a Monte Carlo scheme to sample burst behavior (e.g. amplitude, timescale) from the literature. For a range of ensemble sizes and assumptions of contributions from different burst types, we use the resulting simulated lightcurves to determine if the proposed far-IR observing program with PRIMA would allow for the accurate recovery of burst mass accumulation. The employed program is proposed in PRIMA GO Case \#43 in \citet{Moullet2023} in which five years of PRIMA observations, with an observing cadence of six months, is supplemented by 2009 Herschel data. 

We investigated two distinct cases: one in which the stellar mass assembly process is dominated by a single mode of high-amplitude accretion and one in which multiple outburst types can occur. In case one, we investigated the scenario in which FU Ori outbursts contribute 50\% of a star's total mass using protostellar ensembles ranging in size from 50-5000 protostars. Due to the rarity of FU Ori events, case one acts as a limiting case in which we can determine the minimum protostars an ensemble must have in order for the proposed PRIMA program to accurately constrain the importance of large outbursts in the stellar mass assembly process. Adopting the ensemble size from case one, multiple outburst types were added to the simulation to assess if the proposed program is sufficient in determining the mass contributors in the stellar mass assembly process of each outburst type.

In the scenario in which FU Ori outbursts alone contribute 50\% of a star's total mass, we draw the following conclusions: 

\begin{itemize}
    \item The proposed PRIMA program is able to detect at least one outburst over its observing lifetime in all simulation runs for ensembles of 500-5000 protostars. In ensembles smaller than 500 protostars, no FU Ori outbursts are detected in 50\% of realizations for an ensemble of 50 protostars and in 25\% of realizations for an ensemble of 100 protostars. 
    \item In all ensemble sizes, PRIMA is able to recover $\sim91\%$ of FU Ori outbursts on average when comparing the true and observed number of outbursts, with high degrees of variation in burst recovery percentages for the three smallest ensemble sizes (between 10-30\%). Recovery percentages less then 100\% are expected as we expect to systemically miss outbursts. 
    \item The proposed PRIMA program detects 0.8\% of an ensemble undergoing an outburst for all ensemble sizes, with large probability spreads for ensembles of 50 and 100 protostars. These observed percentages are systematically offset from the true percentage by 0.2\%, an expected. PRIMA will miss five years of brightening activity for each 30 year burst and thus miss 20\% of theses bursts. A significant narrowing of the uncertainty in this probability distribution occurs once an ensemble has at least 500 protostars but narrows further once an ensemble has at least 1000 protostars.
    \item Mass accumulation rates of increasing ensemble sizes tend toward 43\% with ensemble of 50 and 100 protostars showing mass accumulation rates significantly lower than this value, 23\% and 33\%, respectively. These mass accumulation rates align with predictions as mass percentages are a direct result of the number of bursts occurring at any given observation. Increasing ensemble size narrows the probability spread significantly from $\sigma\sim29\%$ for an ensemble of 50 protostars to $\sigma\sim4\%$ for an ensemble of 5000 protostars.
\end{itemize}

Adopting a nominal ensemble size of 2000 protostars from the findings above and varying the number of outburst types allowed to occur, we draw the following conclusions: 

\begin{itemize}
    \item For an ensemble of 2000 protostars, the proposed PRIMA program is able to recover $\sim97-100\%$ of V1647 Ori, EX Lup, and RV-Long outbursts with little spread in this value between realizations. This indicates that burst recovery percentages are robust when outburst duration is within the limits of the program duration and observing cadence. 
    \item FU Ori and RV-Short outbursts have lower burst recovery rates ($\sim89\%$ and $\sim8.5\%$) due to outburst durations being longer than the program lifetime or shorter than the observing cadence, respectively. It is expected that we miss a large amount of RV-Short outbursts and a small amount of FU Ori outbursts as FU Ori outburst durations are marginally longer than the program lifetime, but RV-Short outbursts are significantly shorter than the observing cadence. 
    \item Despite missed outbursts, observed mass accumulation percentages from all outburst types are $\sim10\%$ and are comparable to that of the true mass accumulation rates (offset by a factor between 0.09-2.1\%). Probability distributions for EX Lup, RV-Long, and RV-Short are relatively narrow ($\sigma$ between 0.5-0.8\%), FU Ori and V1647 Ori show wider distributions ($\sigma$ of 3.4\% and 2.1\%, respectively). 
\end{itemize}

Findings indicate that for an ensemble of 2000 protostars in which five outburst types occur, the proposed PRIMA program is able to determine the total mass accumulation percentage with a $1\sigma$ accuracy of 2.3\%. Accuracy per burst type varies with V1647 Ori, EX Lup, RV-Long, and RV-Short having $1\sigma$ accuracies $<2\%$ and FU Ori outbursts having a $1\sigma$ accuracy of $3.4\%$. 

\begin{acknowledgments}
    C.B.  gratefully  acknowledges  funding  from  National  Science  Foundation  under  Award  Nos. 2108938, 2206510, and CAREER 2145689, as well as from the National Aeronautics and Space Administration through the Astrophysics Data Analysis Program under Award ``3-D MC: Mapping Circumnuclear Molecular Clouds from X-ray to Radio,” Grant No. 80NSSC22K1125.
    Y.H. was supported by the Jet Propulsion Laboratory, California Institute of Technology, under a contract with the National Aeronautics and Space Administration (80NM0018D0004). M.S. acknowledges support from the NASA ADAP grant No. 80NSSC22K0168.  The material is based upon work supported by NASA under award number 80GSFC24M0006 (M.S.). 
    D.J. is supported by NRC Canada and by an NSERC Discovery Grant.
    R.L. wishes to thank Jennifer Wallace, Megan Davis, Kaustub Anand, and Eduardo Aguirre Serrata for their support. 
\end{acknowledgments}

\bibliography{reflist}{}

\begin{thebibliography}{}
\expandafter\ifx\csname natexlab\endcsname\relax\def\natexlab#1{#1}\fi
\providecommand{\url}[1]{\href{#1}{#1}}
\providecommand{\dodoi}[1]{doi:~\href{http://doi.org/#1}{\nolinkurl{#1}}}
\providecommand{\doeprint}[1]{\href{http://ascl.net/#1}{\nolinkurl{http://ascl.net/#1}}}
\providecommand{\doarXiv}[1]{\href{https://arxiv.org/abs/#1}{\nolinkurl{https://arxiv.org/abs/#1}}}

\bibitem[{{Acosta-Pulido} {et~al.}(2007){Acosta-Pulido}, {Kun}, {{\'A}brah{\'a}m}, {K{\'o}sp{\'a}l}, {Csizmadia}, {Kiss}, {Mo{\'o}r}, {Szabados}, {Benk{\H{o}}}, {Barrena Delgado}, {Charcos-Llorens}, {Eredics}, {Kiss}, {Manchado}, {R{\'a}cz}, {Ramos Almeida}, {Sz{\'e}kely}, \& {Vidal-N{\'u}{\~n}ez}}]{Acosta-Pulido2007}
{Acosta-Pulido}, J.~A., {Kun}, M., {{\'A}brah{\'a}m}, P., {et~al.} 2007, \aj, 133, 2020, \dodoi{10.1086/512101}

\bibitem[{{Appenzeller}(1972)}]{Appenzeller1972}
{Appenzeller}, I. 1972, Mitteilungen der Astronomischen Gesellschaft Hamburg, 31, 39

\bibitem[{{Aspin} {et~al.}(2010){Aspin}, {Reipurth}, {Herczeg}, \& {Capak}}]{Aspin2010}
{Aspin}, C., {Reipurth}, B., {Herczeg}, G.~J., \& {Capak}, P. 2010, \apjl, 719, L50, \dodoi{10.1088/2041-8205/719/1/L50}

\bibitem[{{Bradford} {et~al.}(2023){Bradford}, {Armus}, {Battersby}, {Bolatto}, {Hensley}, {Kataria}, {Kogut}, {Meixner}, {Mills}, {Moullet}, {Pontoppidan}, {Smith}, {Staguhn}, {Jpl Prima Design Team}, \& {Gsfc Prima Design Team}}]{Bradford2023}
{Bradford}, C., {Armus}, L., {Battersby}, C., {et~al.} 2023, in American Astronomical Society Meeting Abstracts, Vol. 242, American Astronomical Society Meeting Abstracts \#242, 339.02

\bibitem[{{Burgarella} {et~al.}(2023){Burgarella}, {Ciesla}, {Sauvage}, {Prieto}, {Jellema}, {Baselmans}, {Glenn}, {Bradford}, {Pope}, {Armus}, {Battersby}, {Bolatto}, {Hensley}, {Kataria}, {Meixner}, {Mills}, {Moullet}, {Pontoppidan}, {Smith}, {Somerville}, \& {Staguhn}}]{Burgarella2023}
{Burgarella}, D., {Ciesla}, L., {Sauvage}, M., {et~al.} 2023, in American Astronomical Society Meeting Abstracts, Vol. 241, American Astronomical Society Meeting Abstracts \#241, 160.22

\bibitem[{{Caratti o Garatti} {et~al.}(2017){Caratti o Garatti}, {Stecklum}, {Garcia Lopez}, {Eisl{\"o}ffel}, {Ray}, {Sanna}, {Cesaroni}, {Walmsley}, {Oudmaijer}, {de Wit}, {Moscadelli}, {Greiner}, {Krabbe}, {Fischer}, {Klein}, \& {Iba{\~n}ez}}]{CarattioGaratti2017}
{Caratti o Garatti}, A., {Stecklum}, B., {Garcia Lopez}, R., {et~al.} 2017, Nature Physics, 13, 276, \dodoi{10.1038/nphys3942}

\bibitem[{{Cody} {et~al.}(2014){Cody}, {Stauffer}, {Baglin}, {Micela}, {Rebull}, {Flaccomio}, {Morales-Calder{\'o}n}, {Aigrain}, {Bouvier}, {Hillenbrand}, {Gutermuth}, {Song}, {Turner}, {Alencar}, {Zwintz}, {Plavchan}, {Carpenter}, {Findeisen}, {Carey}, {Terebey}, {Hartmann}, {Calvet}, {Teixeira}, {Vrba}, {Wolk}, {Covey}, {Poppenhaeger}, {G{\"u}nther}, {Forbrich}, {Whitney}, {Affer}, {Herbst}, {Hora}, {Barrado}, {Holtzman}, {Marchis}, {Wood}, {Medeiros Guimar{\~a}es}, {Lillo Box}, {Gillen}, {McQuillan}, {Espaillat}, {Allen}, {D'Alessio}, \& {Favata}}]{Cody2014}
{Cody}, A.~M., {Stauffer}, J., {Baglin}, A., {et~al.} 2014, \aj, 147, 82, \dodoi{10.1088/0004-6256/147/4/82}

\bibitem[{{Contreras Pe{\~n}a} {et~al.}(2025){Contreras Pe{\~n}a}, {Lee}, {Lee}, {Herczeg}, {Johnstone}, {Liu}, {Lucas}, {Guo}, {Kuhn}, {Smith}, {Ashraf}, {Jose}, {Yoon}, \& {Yoon}}]{ContrerasPena2025}
{Contreras Pe{\~n}a}, C., {Lee}, J.-E., {Lee}, H.-G., {et~al.} 2025, \apj, 987, 23, \dodoi{10.3847/1538-4357/add25f}

\bibitem[{{Dunham} {et~al.}(2014){Dunham}, {Stutz}, {Allen}, {Evans}, {Fischer}, {Megeath}, {Myers}, {Offner}, {Poteet}, {Tobin}, \& {Vorobyov}}]{Dunham2014}
{Dunham}, M.~M., {Stutz}, A.~M., {Allen}, L.~E., {et~al.} 2014, in Protostars and Planets VI, ed. H.~{Beuther}, R.~S. {Klessen}, C.~P. {Dullemond}, \& T.~{Henning}, 195--218, \dodoi{10.2458/azu_uapress_9780816531240-ch009}

\bibitem[{{Dunham} {et~al.}(2015){Dunham}, {Allen}, {Evans}, {Broekhoven-Fiene}, {Cieza}, {Di Francesco}, {Gutermuth}, {Harvey}, {Hatchell}, {Heiderman}, {Huard}, {Johnstone}, {Kirk}, {Matthews}, {Miller}, {Peterson}, \& {Young}}]{Dunham2015}
{Dunham}, M.~M., {Allen}, L.~E., {Evans}, II, N.~J., {et~al.} 2015, \apjs, 220, 11, \dodoi{10.1088/0067-0049/220/1/11}

\bibitem[{{Evans} {et~al.}(2009){Evans}, {Dunham}, {J{\o}rgensen}, {Enoch}, {Mer{\'\i}n}, {van Dishoeck}, {Alcal{\'a}}, {Myers}, {Stapelfeldt}, {Huard}, {Allen}, {Harvey}, {van Kempen}, {Blake}, {Koerner}, {Mundy}, {Padgett}, \& {Sargent}}]{Evans2009}
{Evans}, II, N.~J., {Dunham}, M.~M., {J{\o}rgensen}, J.~K., {et~al.} 2009, \apjs, 181, 321, \dodoi{10.1088/0067-0049/181/2/321}

\bibitem[{{Fischer} {et~al.}(2023){Fischer}, {Hillenbrand}, {Herczeg}, {Johnstone}, {Kospal}, \& {Dunham}}]{Fischer2023}
{Fischer}, W.~J., {Hillenbrand}, L.~A., {Herczeg}, G.~J., {et~al.} 2023, in Astronomical Society of the Pacific Conference Series, Vol. 534, Protostars and Planets VII, ed. S.~{Inutsuka}, Y.~{Aikawa}, T.~{Muto}, K.~{Tomida}, \& M.~{Tamura}, 355, \dodoi{10.48550/arXiv.2203.11257}

\bibitem[{{Fischer} {et~al.}(2017){Fischer}, {Megeath}, {Furlan}, {Ali}, {Stutz}, {Tobin}, {Osorio}, {Stanke}, {Manoj}, {Poteet}, {Booker}, {Hartmann}, {Wilson}, {Myers}, \& {Watson}}]{Fischer2017}
{Fischer}, W.~J., {Megeath}, S.~T., {Furlan}, E., {et~al.} 2017, \apj, 840, 69, \dodoi{10.3847/1538-4357/aa6d69}

\bibitem[{{Fischer} {et~al.}(2024){Fischer}, {Battersby}, {Johnstone}, {Lee}, {Sewi{\l}o}, {Beuther}, {Hasegawa}, {Ginsburg}, \& {Pontoppidan}}]{Fischer2024}
{Fischer}, W.~J., {Battersby}, C., {Johnstone}, D., {et~al.} 2024, \aj, 167, 82, \dodoi{10.3847/1538-3881/ad188b}

\bibitem[{{Frimann} {et~al.}(2017){Frimann}, {J{\o}rgensen}, {Dunham}, {Bourke}, {Kristensen}, {Offner}, {Stephens}, {Tobin}, \& {Vorobyov}}]{Frimann2017}
{Frimann}, S., {J{\o}rgensen}, J.~K., {Dunham}, M.~M., {et~al.} 2017, \aap, 602, A120, \dodoi{10.1051/0004-6361/201629739}

\bibitem[{{Furlan} {et~al.}(2016){Furlan}, {Fischer}, {Ali}, {Stutz}, {Stanke}, {Tobin}, {Megeath}, {Osorio}, {Hartmann}, {Calvet}, {Poteet}, {Booker}, {Manoj}, {Watson}, \& {Allen}}]{Furlan2016}
{Furlan}, E., {Fischer}, W.~J., {Ali}, B., {et~al.} 2016, \apjs, 224, 5, \dodoi{10.3847/0067-0049/224/1/5}

\bibitem[{{Glenn} {et~al.}(2025){Glenn}, {Meixner}, {Bradford}, {Pontoppidan}, {Pope}, {Kataria}, {Rocca}, {Luthman}, {Armus}, {Baselmans}, {Battersby}, {Bollato}, {Burgarella}, {Chen}, {Ciesla}, {Day}, {Di Giorgio}, {Dipirro}, {Dowell}, {Echternach}, {Essinger-Hileman}, {Foote}, {Gruppioni}, {Hensley}, {Henning}, {Jellema}, {Johnson}, {Kogut}, {Krause}, {McGuire}, {Mills}, {Moullet}, {Rodgers}, {Sauvage}, {Smith}, {Somerville}, {Staguhn}, {Stevenson}, {Tucker}, {Unwin}, {Ziemer}, {Cannella}, \& {Dissly}}]{Glenn2025}
{Glenn}, J., {Meixner}, M., {Bradford}, C.~M., {et~al.} 2025, Journal of Astronomical Telescopes, Instruments, and Systems, 11, 031628, \dodoi{10.1117/1.JATIS.11.3.031628}

\bibitem[{{Gullbring} {et~al.}(1998){Gullbring}, {Hartmann}, {Brice{\~n}o}, \& {Calvet}}]{Gullbring1998}
{Gullbring}, E., {Hartmann}, L., {Brice{\~n}o}, C., \& {Calvet}, N. 1998, \apj, 492, 323, \dodoi{10.1086/305032}

\bibitem[{{Hartmann} \& {Kenyon}(1985)}]{Hartmann1985}
{Hartmann}, L., \& {Kenyon}, S.~J. 1985, \apj, 299, 462, \dodoi{10.1086/163713}

\bibitem[{{Herbig}(1966)}]{Herbig1966}
{Herbig}, G.~H. 1966, Vistas in Astronomy, 8, 109, \dodoi{10.1016/0083-6656(66)90025-0}

\bibitem[{{Herbig}(1989)}]{Herbig1989}
{Herbig}, G.~H. 1989, in European Southern Observatory Conference and Workshop Proceedings, Vol.~33, European Southern Observatory Conference and Workshop Proceedings, ed. B.~{Reipurth}, 233--246

\bibitem[{{Herczeg} {et~al.}(2017){Herczeg}, {Johnstone}, {Mairs}, {Hatchell}, {Lee}, {Bower}, {Chen}, {Aikawa}, {Yoo}, {Kang}, {Kang}, {Chen}, {Williams}, {Bae}, {Dunham}, {Vorobyov}, {Zhu}, {Rao}, {Kirk}, {Takahashi}, {Morata}, {Lacaille}, {Lane}, {Pon}, {Scholz}, {Samal}, {Bell}, {Graves}, {Lee}, {Parsons}, {He}, {Zhou}, {Kim}, {Chapman}, {Drabek-Maunder}, {Chung}, {Eyres}, {Forbrich}, {Hillenbrand}, {Inutsuka}, {Kim}, {Kim}, {Kuan}, {Kwon}, {Lai}, {Lalchand}, {Lee}, {Lee}, {Long}, {Lyo}, {Qian}, {Scicluna}, {Soam}, {Stamatellos}, {Takakuwa}, {Tang}, {Wang}, \& {Wang}}]{Herczeg2017}
{Herczeg}, G.~J., {Johnstone}, D., {Mairs}, S., {et~al.} 2017, \apj, 849, 43, \dodoi{10.3847/1538-4357/aa8b62}

\bibitem[{{J{\o}rgensen} {et~al.}(2020){J{\o}rgensen}, {Belloche}, \& {Garrod}}]{Jorgensen2020}
{J{\o}rgensen}, J.~K., {Belloche}, A., \& {Garrod}, R.~T. 2020, \araa, 58, 727, \dodoi{10.1146/annurev-astro-032620-021927}

\bibitem[{{Kenyon} {et~al.}(1990){Kenyon}, {Hartmann}, {Strom}, \& {Strom}}]{Kenyon1990}
{Kenyon}, S.~J., {Hartmann}, L.~W., {Strom}, K.~M., \& {Strom}, S.~E. 1990, \aj, 99, 869, \dodoi{10.1086/115380}

\bibitem[{{Kim} {et~al.}(2024){Kim}, {Lee}, {Pe{\~n}a}, {Johnstone}, {Herczeg}, {Tobin}, \& {Evans}}]{Kim2024}
{Kim}, C.-H., {Lee}, J.-E., {Pe{\~n}a}, C.~C., {et~al.} 2024, \apj, 961, 108, \dodoi{10.3847/1538-4357/ad1400}

\bibitem[{{Kim} {et~al.}(2012){Kim}, {Evans}, {Dunham}, {Lee}, \& {Pontoppidan}}]{Kim2012}
{Kim}, H.~J., {Evans}, II, N.~J., {Dunham}, M.~M., {Lee}, J.-E., \& {Pontoppidan}, K.~M. 2012, \apj, 758, 38, \dodoi{10.1088/0004-637X/758/1/38}

\bibitem[{{Larson}(1973)}]{Larson1973}
{Larson}, R.~B. 1973, \araa, 11, 219, \dodoi{10.1146/annurev.aa.11.090173.001251}

\bibitem[{{Le Gouellec} {et~al.}(2024){Le Gouellec}, {Greene}, {Hillenbrand}, \& {Yates}}]{LeGouellec2024}
{Le Gouellec}, V. J.~M., {Greene}, T.~P., {Hillenbrand}, L.~A., \& {Yates}, Z. 2024, \apj, 966, 91, \dodoi{10.3847/1538-4357/ad2935}

\bibitem[{{Lee} {et~al.}(2024){Lee}, {Lee}, {Contreras Pe{\~n}a}, {Johnstone}, {Herczeg}, \& {Lee}}]{Lee2024}
{Lee}, S., {Lee}, J.-E., {Contreras Pe{\~n}a}, C., {et~al.} 2024, \apj, 962, 38, \dodoi{10.3847/1538-4357/ad14f8}

\bibitem[{{Mairs} {et~al.}(2024){Mairs}, {Lee}, {Johnstone}, {Broughton}, {Lee}, {Herczeg}, {Bell}, {Chen}, {Contreras-Pe{\~n}a}, {Francis}, {Hatchell}, {Kim}, {Liu}, {Park}, {Qiu}, {Wang}, {Zhang}, \& {JCMT Transient Team}}]{Mairs2024}
{Mairs}, S., {Lee}, S., {Johnstone}, D., {et~al.} 2024, \apj, 966, 215, \dodoi{10.3847/1538-4357/ad35b6}

\bibitem[{{Moullet} {et~al.}(2023){Moullet}, {Kataria}, {Lis}, {Unwin}, {Hasegawa}, {Mills}, {Battersby}, {Roc}, \& {Meixner}}]{Moullet2023}
{Moullet}, A., {Kataria}, T., {Lis}, D., {et~al.} 2023, arXiv e-prints, arXiv:2310.20572, \dodoi{10.48550/arXiv.2310.20572}

\bibitem[{{Ninan} {et~al.}(2013){Ninan}, {Ojha}, {Bhatt}, {Ghosh}, {Mohan}, {Mallick}, {Tamura}, \& {Henning}}]{Ninan2013}
{Ninan}, J.~P., {Ojha}, D.~K., {Bhatt}, B.~C., {et~al.} 2013, \apj, 778, 116, \dodoi{10.1088/0004-637X/778/2/116}

\bibitem[{{Park} {et~al.}(2021){Park}, {Lee}, {Contreras Pe{\~n}a}, {Johnstone}, {Herczeg}, {Lee}, {Lee}, {Bhardwaj}, \& {Moriarty-Schieven}}]{Park2021}
{Park}, W., {Lee}, J.-E., {Contreras Pe{\~n}a}, C., {et~al.} 2021, \apj, 920, 132, \dodoi{10.3847/1538-4357/ac1745}

\bibitem[{{Pilbratt} {et~al.}(2010){Pilbratt}, {Riedinger}, {Passvogel}, {Crone}, {Doyle}, {Gageur}, {Heras}, {Jewell}, {Metcalfe}, {Ott}, \& {Schmidt}}]{Pilbratt2010}
{Pilbratt}, G.~L., {Riedinger}, J.~R., {Passvogel}, T., {et~al.} 2010, \aap, 518, L1, \dodoi{10.1051/0004-6361/201014759}

\bibitem[{{Plunkett} {et~al.}(2015){Plunkett}, {Arce}, {Mardones}, {van Dokkum}, {Dunham}, {Fern{\'a}ndez-L{\'o}pez}, {Gallardo}, \& {Corder}}]{Plunkett2015}
{Plunkett}, A.~L., {Arce}, H.~G., {Mardones}, D., {et~al.} 2015, \nat, 527, 70, \dodoi{10.1038/nature15702}

\bibitem[{{Ray} {et~al.}(2023){Ray}, {McCaughrean}, {Caratti o Garatti}, {Kavanagh}, {Justtanont}, {van Dishoeck}, {Reitsma}, {Beuther}, {Francis}, {Gieser}, {Klaassen}, {Perotti}, {Tychoniec}, {van Gelder}, {Colina}, {Greve}, {G{\"u}del}, {Henning}, {Lagage}, {{\"O}stlin}, {Vandenbussche}, {Waelkens}, \& {Wright}}]{Ray2023}
{Ray}, T.~P., {McCaughrean}, M.~J., {Caratti o Garatti}, A., {et~al.} 2023, \nat, 622, 48, \dodoi{10.1038/s41586-023-06551-1}

\bibitem[{{Reipurth}(1989)}]{Reipurth1989}
{Reipurth}, B. 1989, \nat, 340, 42, \dodoi{10.1038/340042a0}

\bibitem[{{Shu}(1977)}]{Shu1977}
{Shu}, F.~H. 1977, \apj, 214, 488, \dodoi{10.1086/155274}

\bibitem[{{Shu} {et~al.}(1987){Shu}, {Lizano}, \& {Adams}}]{Shu1987}
{Shu}, F.~H., {Lizano}, S., \& {Adams}, F.~C. 1987, in IAU Symposium, Vol. 115, Star Forming Regions, ed. M.~{Peimbert} \& J.~{Jugaku}, 417--433

\end{thebibliography}
\bibliographystyle{aasjournal}

\section{Appendix}\label{sec:appendix}

\subsection{Example Calculation - Case One}\label{sec:calculation_case_one}
Here we show an example calculation of how our burst probabilities are determined. For this example, we look at the burst probability of a FU Ori burst happening in a one burst type scenario, i.e. the FU Ori bursts contributing 50\% of the overall mass of the star and 100\% of the episodic mass. 

In this case, N=1 (FU Ori only) and $f_{i,m}$ = 50\%. This means Equation~\ref{eq:burst_mass_frac_duration} can be written as:

\begin{equation}{\label{eq:one_bin_FU_mass}}
    f_{\rm FU,m} = \frac{\dot{M}_{\rm SS}A_{\rm FU}\tau_{\rm FU}}{\dot{M}_{\rm SS}\tau_{\rm SS} + \dot{M}_{\rm SS}A_{\rm FU}\tau_{\rm FU}}
\end{equation}

Since we do not know the protostellar lifetime, the time spent in steady state, or the time spent in outburst, we rewrite these timescales in terms of the fractional times:

\begin{eqnarray} \label{eq:fractional_times1}
    \frac{\tau_{\rm FU}}{\tau_{\rm lt}} & = & \frac{\tau_{\rm FU}}{\tau_{\rm SS}+\tau_{\rm FU}} = f_{\rm FU,t}\\ 
    \frac{\tau_{\rm SS}}{\tau_{\rm lt}} & = & \frac{\tau_{\rm SS}}{\tau_{\rm SS}+\tau_{\rm FU}} = 1-f_{\rm FU,t}
\end{eqnarray}

Plugging these into Equation~\ref{eq:one_bin_FU_mass}:

\begin{equation}{\label{eq:one_bin_FU_mass_withLT}}
    f_{\rm FU,m} = \frac{\dot{M}_{\rm SS}A_{\rm FU}f_{\rm FU,t}\tau_{\rm lt}}{\dot{M}_{\rm SS}(1-f_{\rm FU,t})\tau_{\rm lt} + \dot{M}_{\rm SS}A_{\rm FU}f_{\rm  FU,t}\tau_{\rm lt}} = \frac{A_{\rm FU}f_{\rm FU,t}}{(1-f_{\rm FU,t}) + A_{\rm FU}f_{\rm FU,t}}
\end{equation}

Plug in A=100 and $f_{\rm FU,m}$=50\% and solve for $f_{\rm FU,t}$:
\begin{eqnarray} \label{eq:solve_f_it}
    0.5 &=& \frac{100f_{\rm FU,t}}{(1-f_{\rm FU,t}) + 100f_{\rm FU,t}}\\
    0.5 &=& \frac{100f_{\rm FU,t}}{1+99f_{\rm FU,t}}\\
    1.0 &=& 101f_{\rm FU,t}\\
    f_{\rm FU,t} &=& \frac{1}{101} = 0.0099
\end{eqnarray}

This implies that a star must spend $\sim$1\% of its lifetime in an FU Ori outburst. Dividing this value by the outburst duration as in Equation~\ref{eq:burst_prob} and assuming a duration of 30 years, we recover a burst probability of $\mathbf{p_{\rm FU} = 3.30 \cdot 10^{-4}}$ bursts per year. 

\subsection{Example Calculation - Case Two}\label{sec:calculation_case_two}
Here we show the calculation of how our burst probabilities are determined in case two of our simulations in which all burst (FU Ori, V1647 Ori, EX Lup, RV-Long, and RV-Short) types may occur and each type of burst contributes 10\% of the total final mass of the star.

In this case, N=5 and $f_{i,m}$ = 10\%. This means Equation 6 can be written as:

\begin{equation}{\label{eq:five_bin_i_mass}}
    f_{i,m} = \frac{\dot{M}_{\rm SS}A_{i}\tau_{i}}{\dot{M}_{\rm SS}\left(\tau_{\rm SS} + A_{\rm FU}\tau_{\rm FU} + A_{\rm V1647}\tau_{\rm V1647} + A_{\rm EX}\tau_{\rm EX} + A_{\rm RVL}\tau_{\rm RVL} + A_{\rm RVS}\tau_{\rm RVS}\right)},
\end{equation}

Since we do not know the protostellar lifetime, the time spent in steady state, or the time spent in outburst, we rewrite these timescales in terms of the fractional times:

\begin{eqnarray} \label{eq:fractional_times2}
    \frac{\tau_{\rm FU}}{\tau_{\rm lt}} & = & \frac{\tau_{\rm FU}}{\tau_{\rm SS}+\tau_{\rm FU} + \tau_{\rm V1647}+ \tau_{\rm EX}+ \tau_{\rm RVL} + \tau_{\rm RVS}} = f_{\rm FU,t}\\ 
    \frac{\tau_{\rm V1647}}{\tau_{\rm lt}} & = & \frac{\tau_{\rm V1647}}{\tau_{\rm SS}+\tau_{\rm FU} + \tau_{\rm V1647}+ \tau_{\rm EX}+ \tau_{\rm RVL} + \tau_{\rm RVS}} = f_{\rm V1647,t}\\ 
    \frac{\tau_{\rm EX}}{\tau_{\rm lt}} & = & \frac{\tau_{\rm EX}}{\tau_{\rm SS}+\tau_{\rm FU} + \tau_{\rm V1647}+ \tau_{\rm EX}+ \tau_{\rm RVL} + \tau_{\rm RVS}} = f_{\rm EX,t}\\ 
    \frac{\tau_{\rm RVL}}{\tau_{\rm lt}} & = & \frac{\tau_{\rm RVL}}{\tau_{\rm SS}+\tau_{\rm FU} + \tau_{\rm V1647}+ \tau_{\rm EX}+ \tau_{\rm RVL} + \tau_{\rm RVS}} = f_{\rm RVL,t}\\ 
    \frac{\tau_{\rm RVS}}{\tau_{\rm lt}} & = & \frac{\tau_{\rm RVS}}{\tau_{\rm SS}+\tau_{\rm FU} + \tau_{\rm V1647}+ \tau_{\rm EX}+ \tau_{\rm RVL} + \tau_{\rm RVS}} = f_{\rm RVS,t}\\ 
    \frac{\tau_{\rm SS}}{\tau_{\rm lt}} & = & 1-f_{\rm FU,t}-f_{\rm V1647,t}-f_{\rm EX,t}-f_{\rm RVL,t}-f_{\rm RVS,t}
\end{eqnarray}

Plugging these into Equation~\ref{eq:five_bin_i_mass} and considering FU Ori outbursts first: 

\begin{equation}{\label{eq:five_bin_FU_mass_withLT}}
    f_{\rm FU,m} = \frac{A_{\rm FU}f_{\rm FU,t}\tau_{\rm lt}}{(1-\Sigma^{N}_{j}f_{j,t})\tau_{\rm lt} +\Sigma^{N}_{j}f_{j,t}\tau_{\rm lt}} = \frac{A_{\rm FU}f_{\rm FU,t}}{(1-\Sigma^{N}_{j}f_{j,t}) + \Sigma^{N}_{j}f_{j,t}},
\end{equation}

Where $\Sigma^{N}_{j}f_{j,t} = f_{\rm FU,t} + f_{\rm V1647,t} + f_{\rm EX,t} + f_{\rm RVL,t} + f_{\rm RVS,t}$. Plugging in $A_{\rm FU} = 100$, $A_{\rm V1647} = 40$, $A_{\rm EX} = 16$, $A_{\rm RVL} = 6$, $A_{\rm RVS} = 3$ and $f_{\rm FU,m}$=10\%, we solve for $f_{\rm FU,t}$:

\begin{eqnarray} \label{eq:solve_f_it5}
    0.1 &=& \frac{100f_{\rm FU,t}}{1+99f_{\rm FU,t}+39f_{\rm V1467,t}+15f_{\rm EX,t}+5f_{\rm RVL,t}+2f_{\rm RVS,t}}\\
    1.0 &=& 901f_{\rm FU,t} - 39f_{\rm V1647,t} - 15f_{\rm EX,t} - 5f_{\rm RVL,t} - 2f_{\rm RVS,t}
\end{eqnarray}

Repeating these steps for the other burst types, we get the following system of equations: 

\begin{eqnarray} \label{eq:solve_f_it5_all}
    1.0 &=& 901f_{\rm FU,t} - 39f_{\rm V1647,t} - 15f_{\rm EX,t} - 5f_{\rm RVL,t} - 2f_{\rm RVS,t}\\
    1.0 &=& -99f_{\rm FU,t} + 361f_{\rm V1647,t} - 15f_{\rm EX,t} - 5f_{\rm RVL,t} - 2f_{\rm RVS,t}\\
    1.0 &=& -99f_{\rm FU,t} - 39f_{\rm V1647,t} + 145f_{\rm EX,t} - 5f_{\rm RVL,t} - 2f_{\rm RVS,t}\\
    1.0 &=& -99f_{\rm FU,t} - 39f_{\rm V1647,t} - 15f_{\rm EX,t} + 55f_{\rm RVL,t} - 2f_{\rm RVS,t}\\
    1.0 &=& -99f_{\rm FU,t} - 39f_{\rm V1647,t} - 15f_{\rm EX,t} - 5f_{\rm RVL,t} + 28f_{\rm RVS,t}
\end{eqnarray}

Solving these gives the following fractions of time spent in outburst:

\begin{eqnarray} \label{eq:values_f_it5_all}
    f_{\rm FU,t} &=& 1.79 \cdot 10^{-3}\\
    f_{\rm V1647,t} &=& 4.47 \cdot 10^{-3}\\
    f_{\rm EX,t} &=& 1.12 \cdot 10^{-2}\\
    f_{\rm RVL,t} &=& 2.98 \cdot 10^{-2}\\
    f_{\rm RVS,t} &=& 5.96 \cdot 10^{-2}
\end{eqnarray}

Using Equation~\ref{eq:burst_prob} and assuming $t_{\rm FU} = 30$yr, $t_{\rm V1647} = 3$yr, $t_{\rm EX} = 1$yr, $t_{\rm RVL} = 3$yr, $t_{\rm RVS} = 0.04$yr, probabilities for each burst type are:

\begin{eqnarray} \label{eq:values_prob5_all}
    f_{\rm FU,t} &=& 5.96 \cdot 10^{-5}\ [\rm burst\ yr^{-1}]\\
    f_{\rm V1647,t} &=& 1.49 \cdot 10^{-3}\ [\rm burst\ yr^{-1}]\\
    f_{\rm EX,t} &=& 0.0111\ [\rm burst\ yr^{-1}]\\
    f_{\rm RVL,t} &=& 1.4889\ [\rm burst\ yr^{-1}]\\
    f_{\rm RVS,t} &=& 0.0099\ [\rm burst\ yr^{-1}]
\end{eqnarray}

\end{document}